\documentstyle[11pt]{article}
\def\a{\alpha}

\def\rar{\rightarrow}

\def\le{\left(}
\def\ri{\right)}
\def\t{\theta}

\def\ve{\varepsilon}

\def\f12{\frac{1}{2}}
\def\fra1g2{\frac{1}{g^2}}

\def\dis{\displaystyle}
\def\del{\delta}
\def\G{\Gamma}

\def\no{\nonumber}

\def\pd{\partial}

\begin{document}
\begin{titlepage}
\flushright{Brown-HET-1408}\\
\flushright{USM-TH-158}\\
\vskip 2cm
\begin{center}
{\Large \bf Semiclassical scattering amplitudes of dressed gravitons} \\
\vskip 1cm  
Kyungsik Kang$^{(a)}$ and Igor Kondrashuk$^{(b)}$ 
\vskip 5mm 
{\it (a) Physics Department, Brown University, Providence, Rhode Island 02912, USA} \\
\vskip 5mm  
{\it (b) Departamento de F\'\i sica, 
Universidad T\'ecnica Federico Santa Mar\'\i a, \\ 
Avenida Espa\~{n}a 1680, Casilla 110-V, Valparaiso, Chile} \\
\vskip 20mm
\begin{abstract}
We consider effective action for the Einstein gravity and show that 
dressed mean fields are actual variables of the effective action. 
Kernels of this effective action expressed in terms of dressed effective fields
are constituent parts of scattering amplitudes for  
gravitons. Possible applications to the graviton scattering and 
black hole formation are discussed 
at the semiclassical level. In particular, we consider graviton 
scattering in four dimensions based on the Lipatov effective 
action for quantum gravity, shock waves of particles moving on the brane 
in Randall-Sundrum scenario with fifth extra dimension, and Giddings' 
estimation of Froissart bound. 
\end{abstract}
\end{center}
\end{titlepage}

\section{Introduction}

There is an intriguing belief that the high energy scattering  in gauge theories 
in four space-time dimensions can be described by the bulk physics of 
supergravity with a higher dimension in AdS space. In other words, 
one can extract information about quantum process 
amplitudes in four spacetime dimensions starting with the classical physics 
of wider theories including gravity in higher dimensions. This is called 
gauge/string-gravity duality and is based on Maldacena conjecture 
\cite{Maldacena:1997re}.

The total cross section for the particle absorption by a
black hole is estimated by the area of the black hole horizon in 
Ref. \cite{Das} for four dimensions in a semiclassical way. Another 
estimation has been given in Ref.\cite{Solodukhin:2002ui}. 
In Ref. \cite{Matschull:1998rv} it has been shown that the horizon is necessarily 
formed in three dimensions at sufficiently high energies when two particles 
collide. The amplitude of this process has been calculated at the 
semiclassical level in Ref. \cite{Jevicki:2002fq}, in which classical Hamiltonian 
can be found.

In four dimensions there are no exact results about 
the cross section of the process {\em two particles $\rar$ black hole} 
even at the classical level. One has to rely on approximate estimates 
such as in  Ref. \cite{Das,Solodukhin:2002ui}. In unrestricted four-dimensional 
spacetime the horizon area  
is proportional to the square of total energy in center-mass 
frame. However, in brane models as it has been shown by 
Giddings \cite{Giddings} 
radius of horizon does modify its form to logarithm of c.m. energy. 
This is the basic source of the Froissart bound (FB) for the total cross sections 
which has been known since 1961 as a consequence of unitarity, positivity of 
the imaginary part of partial wave amplitudes of analytic S-matrix 
\cite{Froissart}. Giddings \cite{Giddings} concluded that bulk theory 
feels in some way unitarity 
of the four-dimensional theory on the boundary.

At the same time there is approach to consider the effective action of 
quantum gravity as a functional of the effective field of metric 
convoluted with unspecified dressing function 
\cite{Kondrashuk:2000br,Cvetic:2002dx,Cvetic:2002in,Kondrashuk:2003tw}. 
We will call this construction ``dressed 
effective field''. In this paper we make steps to 
embed the results of semiclassical and eikonal estimations of the 
amplitudes {\em two particles $\rar$ black hole} process and graviton 
scattering processes into the approach 
of Ref. \cite{Cvetic:2002dx,Kondrashuk:2003tw,Cvetic:2004kx}.

The effective action for quantum gravity is Legendre transform of 
the logarithm of path integral. Tilded fields in this paper will mean
dressed effective fields  $\tilde{h}$ and $\tilde{\phi}$ of graviton 
and matter fields, that is the effective fields convoluted 
with unspecified dressing function
\cite{Kondrashuk:2000br,Cvetic:2002dx,Cvetic:2002in,Kondrashuk:2003tw},
\begin{eqnarray*} 
& \dis{ \tilde{h}_{\mu\nu}(x) = \int~d^D x G^{-1}_{h}(x-x')h_{\mu\nu}(x')} \\ 
& \dis{\tilde{\phi}(x) = \int~d^D x  G^{-1}_{\phi}(x-x') \phi(x') },
\end{eqnarray*} 
$h_{\mu\nu}$ is the graviton tensor, $g_{\mu\nu} = \eta_{\mu\nu} + 
h_{\mu\nu},$ where $\eta_{\mu\nu}$ is Minkowski tensor.

In addition to analyzing size of horizon along the lines of Ref. \cite{Giddings}
we consider graviton scattering in four dimensions based on the Lipatov effective 
action for quantum gravity \cite{Lipatov:1991nf}, scattering of one particle by a 
shock wave of another one both of which move on on the brane in Randall-Sundrum 
scenario with fifth extra dimension. We start our analysis with introducing 
the concept of the dressed effective fields.

\section{Effective action of dressed gravitons in four dimensions} 
\label{s2}

There is a way to write a functional structure of the 
effective action for dressed gravitons in  supergravity 
\cite{Kondrashuk:2003tw}. 
For example, one can consider supergravity in the component formulation. 
Strictly speaking, if one works with supergravity the vielbeins must be 
introduced. Since we do not want to overload notation in this paper 
we work with the metric as an independent field which is variable 
of integration in the path integral. This 
would be sufficient for the purpose of the present paper. To clarify the 
idea we elaborate usual four-dimensional Einstein gravity as an example 
in this section.

Let us take the path integral for 
the theory of four dimensional gravity in the following form 
\cite{Hawking'76}:
\begin{eqnarray*}
Z = \int~dg~d{\phi}~\exp i\left[\dis{S[g,\phi]} \right],
\end{eqnarray*} 
where $S[g,\phi]$ is the classical action of D4 gravity $g$ coupled to matter
fields $\phi.$ The action of the gravitational field is usually taken to be 
\begin{eqnarray*}
S = \frac{1}{16\pi}\int~R\sqrt{-g}~d^4x. 
\end{eqnarray*}

In such approach we can fix the symmetry of the diffeomorphism 
group by imposing some linear gauge fixing condition on the graviton
field $g_{\mu\nu}.$ One can take the most general form of a linear 
gauge fixing condition 
\begin{eqnarray*}
F[g_{\mu\nu}] = 0. 
\end{eqnarray*}   
To fix the gauge we have to introduce into the path integral the gauge fixing
term and Faddeev--Popov ghost field at the same time in order to factorize the 
volume of diffeomorphism out \cite{SF}. This procedure brings out additional 
symmetry for the classical action extended by gauge fixing term and by the 
ghost term which is called BRST symmetry \cite{BRST}. Total action including 
gauge fixing, FP ghost action,  at the classical level 
can be written as      
\begin{eqnarray}
& \dis{S = S_{\rm cl} + S_{\rm gf} + S_{\rm gh}} \no\\ 
& = \dis{\frac{1}{16\pi}\int~d^4x~R\sqrt{-g}
- \int~d^4x ~\frac{1}{2\a}\le F[g_{\mu\nu}] \ri^2} \no\\
&  - \dis{\int~d^4x~i~b~\frac{\del F}{\del g_{\mu\nu}}~{\cal L}_c g_{\mu\nu} 
+ S_{I}\le g_{\mu\nu},\phi\ri,}
  \label{total-action-gravity}
\end{eqnarray}
where $c^{\mu}(x)$ here is ghost field, $b(x)$ is antighost field,  
and ${\cal L}_c$ is Lie derivative associated with ghost field 
$c ^{\mu}(x)$ that acts on the metric field as
\begin{eqnarray*}
{\cal L}_c g^{\mu\nu} = c^\lambda\pd_\lambda g^{\mu\nu}- (\pd_\lambda c^\mu)
g^{\nu\lambda} - (\pd_\lambda c^\nu)g^{\mu\lambda},
\end{eqnarray*}
and $S_{I} (g, \phi)$ is the action term containing the interaction 
between gravity and matter. The BRST symmetry for the action 
(\ref{total-action-gravity}) can be written as  
\begin{eqnarray}
& \dis{g_{\mu\nu}  \rar g_{\mu\nu} + i{\cal L}_c g_{\mu\nu} \ve },  \no \\
& \dis{\phi \rar \phi + i{\cal L}_c \phi \ve },  \no \\
& \dis{c \rar c - \frac{1}{2}{\cal L}_c{c}\ve}, \no \\
& \dis{b \rar b + \frac{1}{\a}F\ve,}  
\label{BRST}
\end{eqnarray}   
$\ve$ is Grassmannian constant parameter of the BRST symmetry, $\ve^2= 0.$ 
This invariance at the level of quantum theory can be transformed 
in a usual way \cite{SF} to Slavnov--Taylor (ST) identity that is the 
equation for the Legendre transform of the logarithm of the 
path integral. This Legendre transform is performed 
with respect to external sources of the theory which are 
coupled in the path integral to the quantum fields from the measure of 
the path integral. To do this one defines 
the path integral extended by the dependence on the following external 
sources 
\begin{eqnarray}
& \dis{Z[T,~\eta,~\rho,~K,~k,~L] 
= \int~dg_{\mu\nu}~dc_{\lambda}~db~d\phi~
 \exp i}\left[\dis{S} \right.  \no \\
& \left. + \dis{\int~d^4x~T^{\mu\nu}g_{\mu\nu} + \int~d^4x~j~\phi 
+ i\int~d^4x~\eta_{\mu}c^{\mu}
+ i\int~d^4x~\rho b} \right. \label{path-gravity}\\
& \left. + ~\dis{i\int~d^4x~K^{\mu\nu}{\cal L}_c g_{\mu\nu}  +
 i\int~d^4x~k{\cal L}_c \phi - 
\int~d^4x~L_\mu \frac{1}{2} {\cal L}_c{c^\mu}}\right], 
\no
\end{eqnarray}
where new external sources $K^{\mu\nu},$  $k,$ and $L_\mu$ coupled 
to the BRST variations of the metric, matter field and the ghost field
under group of diffeomorphisms are introduced, respectively. 
The effective action $\G$ is related to $W = i~ln~Z$ by the Legendre
transformation
\begin{eqnarray}
& \dis{g_{\mu\nu}  \equiv - \frac{\del W}{\del T^{\mu\nu}},~~~
\phi \equiv - \frac{\del W}{\del j}, 
  ~~~ ic^\mu  \equiv - \frac{\del W}{\del \eta_\mu}, ~~
  ib \equiv - \frac{\del W}{\del \rho}}, ~~\label{defphi}\\
& \dis{\G = - W - \int~d^4x~T^{\mu\nu}g_{\mu\nu}
- \int~d^4x~j~\phi 
- i\int~d^4x~\eta_\mu c^\mu - i\int~d^4x~\rho b}\no 
\end{eqnarray}
If all equations Eq.~(\ref{defphi}) can be inverted,
\begin{eqnarray*}
& \Omega = \Omega[\varphi,K^{\mu\nu},k,L_\mu], \\
& \dis{\Omega \equiv \le T^{\mu\nu},j,\eta_\mu,\rho\ri,  ~~~   
\varphi \equiv \le g_{\mu\nu}, \phi, c^\mu, b\ri}. \no
\end{eqnarray*}
the effective action can be  defined in terms of new variables,
$\G = \G[\varphi,K^{\mu\nu},k,L_\mu].$
Hence the following equalities hold:
\begin{eqnarray}
& \dis{\frac{\del \G}{\del g_{\mu\nu} } = - T^{\mu\nu} , ~~~ 
 \frac{\del \G}{\del \phi} =  - j, ~~~
\frac{\del \G}{\del K^{\mu\nu}} = - \frac{\del W}{\del K^{\mu\nu}}}, 
\label{GW}\\
& \dis{\frac{\del \G}{\del k} 
= - \frac{\del W}{\del k}, ~~~
\frac{\del \G}{\del c^\mu} = i\eta_\mu, ~~ \frac{\del \G}{\del b} = 
i\rho,  ~~ \frac{\del \G}{\del L_\mu} = - \frac{\del W}{\del L_\mu}}. \no
\end{eqnarray}

If the transformation Eq.~(\ref{BRST}) is made in the path integral 
Eq.~(\ref{path-gravity}) one obtains (as the result of the invariance of the
integral Eq.~(\ref{path-gravity}) under a change of variables) the 
ST identity:
\begin{eqnarray*}
& \dis{\left[\int~d^4x~T^{\mu\nu}\frac{\del}{\del K^{\mu\nu}}
+ \int~d^4x j\frac{\del }{\del k}
- \int~d^4x~i\eta_\mu\le\frac{\del}{\del L_\mu}\ri \right.} \no\\
& + \dis{\left. 
\int~d^4x~i\rho\frac{1}{\a}F\left[\frac{\del}{\del T^{\mu\nu}}\right]
\right]W} = 0,
\end{eqnarray*}
or, taking into account the relations Eq.~(\ref{GW}), we have
\begin{eqnarray}
& \dis{\int~d^4x~\frac{\del \G}{\del g_{\mu\nu}}\frac{\del \G}{\del K^{\mu\nu}} + 
\int~d^4x~\frac{\del \G}{\del  \phi}
\frac{\del \G}{\del k}
+ \int~d^4x~\frac{\del \G}{\del c^\mu}\frac{\del \G}{\del L_\mu}} \no\\
& \dis{- \int~d^4x~\frac{\del \G}{\del b}\le\frac{1}{\a}F[g_{\mu\nu}]\ri}
 = 0.
\label{STrM}
\end{eqnarray}
In addition to ST identity also there is the ghost equation
that can be derived by shifting the antighost field $b$ by an arbitrary 
field $\ve(x)$ in the path integral. The consequence of 
invariance of the path integral with respect to such a change of 
variable is (in terms of the variables (\ref{defphi})) \cite{SF}
\begin{eqnarray}
\frac{\del \G}{\del b} + \frac{\del F}{\del g_{\mu\nu}}
\frac{\del \G}{\del K^{\mu\nu} } = 0.
 \label{ghost}
\end{eqnarray}
The ghost equation (\ref{ghost}) restricts 
the dependence of $\G$ on the antighost field $b$ and  the external 
source $K_M$ to an arbitrary dependence on the combination 
\begin{eqnarray}
\frac{\del F}{\del g_{\mu\nu}} b(x) -  K^{\mu\nu}(x). \label{comb}
\end{eqnarray}
The main idea of  Refs. 
\cite{Cvetic:2002dx,Cvetic:2002in}
is that the momentum-dependent part of the $Lcc$ correlator related to the 
superficial divergence (divergent in the limit of removing regularization)
is invariant itself with respect to ST identity in each order of the 
perturbation theory. According to Ref. 
\cite{Cvetic:2002dx,Cvetic:2004kx}, this invariance results in the following 
integral equation for the part of the correlator $Lcc$ corresponding  to the 
superficial divergence $\sim \ln\frac{p^2}{\Lambda^2}$  
\begin{eqnarray}
& \dis{\int~dx~\G_{\Lambda}(y',x,z')\G_{\Lambda} (x,y,z) = \int~dx~\G_{\Lambda} 
(y',y,x)\G_{\Lambda} (x,z,z')} \no\\
& = \dis{\int~dx~ \G_{\Lambda} (y',x,z)\G _{\Lambda} (x,z',y),}  \label{integral}
\end{eqnarray}
where $\G_{\Lambda} (x,y,z)$ is the scale-dependent part of the correlator $Lcc$
corresponding to the superficial divergence in each order of the perturbation theory
\cite{Cvetic:2004kx}, $\Lambda$ is a scale of ultraviolet regularization. 
The most general parametrization of this correlator is $\G (x,y,z)$ 
\footnote{The structure of indices in (\ref{A1}) is appropriate for the part 
of the correlator $Lcc$ related to its superficial divergence at least.},
\begin{eqnarray}
\dis{\G \sim \int~dx~dy~dz~\G (x,y,z) 
L_\mu(x)c^\lambda (y) \pd_\lambda c^{\mu}}. \label{A1}
\end{eqnarray}
As has been shown in Ref. \cite{Cvetic:2002dx}, the only solution to the 
integral equation (\ref{integral}) is 
\begin{eqnarray}
& \dis{\G_{\Lambda}(x,y,z) = \int~dx'~G_{c}(x'-x)~G^{-1}_{c}(x'-y)
     ~G^{-1}_{c}(x'-z),}        \label{result}
\end{eqnarray}
where $G_c(x)$ is some unspecified function, $\int d^4x' G_c(x-x')G^{-1}_c(x'-y) = 
\del(x-y).$ The complete correlator $Lcc$ can be then written as  
\begin{eqnarray*}
& \dis{\int~dx~dy~dz~\G (x,y,z) L_\mu(x)
c^\lambda (y) \pd_\lambda c^{\mu}(z) } = \\
& = \dis{\int~dx'dy'dz'dxdydz~\tilde{\G} (x',y',z')G_{c}(x'-x)~G^{-1}_{c}(y'-y)}
\times \no\\
& \times~\dis{G^{-1}_{c}(z'-z) L_\mu(x)
c^\lambda (y) \pd_\lambda c^{\mu}(z)} \no\\
& = \dis{\int~dx'dy'dz'~\tilde{\G} (x',y',z')
\tilde{L}_\mu(x')\tilde{c}^\lambda (y') \pd_\lambda \tilde{c}^{\mu}(z')}.
\end{eqnarray*}
Here $\tilde{\G} (x',y',z')$ is the kernel of $Lcc$ correlator written in terms of
dressed fields $\tilde{L}_\mu$  and  $\tilde{c}^\mu$ defined by the convolutions 
\begin{eqnarray*} 
& \dis{ \tilde{L}_{\mu}(x) = \int~d^4 x' G_{c}(x-x')L_\mu(x'),} \\ 
& \dis{ \tilde{c}_{\mu}(x) = \int~d^4 x' G^{-1}_{c}(x-x')c_\mu(x').}
\end{eqnarray*}

As has been shown in Ref. \cite{Cvetic:2004kx}, the consequence of such a 
structure for the $Lcc$ correlator is that the effective action can be expressed 
in terms of the dressed effective fields for the rest of proper correlators.  
The effective action expressed in terms of the dressed effective fields 
possesses some 
kernels which in fact are constituent parts of the scattering 
amplitudes. In ${\cal N} = 4$ super-Yang--Mills theory such kernels 
do not depend on the ultraviolet regularization scale \cite{Cvetic:2004kx}. 
It is known 
for a long time that gravity cannot be renormalized by adding 
finite number of counterterms to remove all the divergences. One can 
think that gravity is regularized in some way that conserves symmetry 
of the classical action, for example by dimensional regularization.
These kernels will depend on the UV scale and on
mutual distances of $n$ spacetime coordinates  of some $n$-point 
proper Green function. For scattering amplitudes of gravitons in multi-Regge 
kinematic region of the momentum space these kernels 
can be calculated in the leading logarithmic approximations as it 
has been shown in Ref.  \cite{Lipatov:1991nf}.  We will use this 
action in the next section.

The structure of background metric in quantum theory of gravity is open question
\cite{Hawking'96}. To analyze the structure of the background metric would 
be possible if we knew physical part of the effective action exactly. 
The minimum of the physical part determines the background metric.  
The background is to be found by solving the equation of motion 
for the effective action in terms of the dressed effective fields of 
matter sector and gravity sector. We will assume that the solution to such 
equations of motion gives
Minkowski or AdS metric on the gravity side. 
 However, metric perturbations about 
the background geometry in course of particle collisions 
can become strong and horizon forms.

\section{Graviton scattering in four dimensions}

Based on the construction of the previous section,  let us begin with the
pure four-dimensional gravity without any higher dimensional
additions. Typical examples considered in literature  concern collision of two
shock waves  in Aichelburg-Sexl  gauge \cite{Aichelburg:1970dh}. There are  
two physically different cases,  both of which  have been
analyzed  in references  at  the semiclassical level.  First  case is  the
scattering of one particle by a  shock wave of another. Second case is
gravitational collapse of two  particles into black hole. The effective
parameter to  differentiate  these two  cases  is
fraction of the effective Schwartzschild radius to the impact parameter of
the problem. In the first case the black hole does not appear.

Lipatov  \cite{Lipatov:1991nf}  has   proposed  to  calculate 
scattering amplitudes of gravitons from some  effective action which 
is local. This
action    restores    unitarity    which    is   lost    in  the leading
approximation.  According  to  Ref.  \cite{Amati:1993tb}  the  eikonal
amplitude can be  calculated by solving equations of  motion that come
from the  extremizing the Lipatov  action and estimating the  value of
this action on the effective trajectories of motion. The action of 
Ref. \cite{Lipatov:1991nf} gives the same results for the tree level 
scattering amplitudes of gravitons. In Ref.  \cite{Amati:1993tb} 
it has been checked that the Lipatov action reproduces eikonal amplitudes 
correctly.

The Lipatov effective action  \cite{Lipatov:1991nf} is 
\begin{eqnarray}
\int d^4x~ L_{eff} = \int d^4x~\le L_0 + L_e +  L_r \ri, 
\end{eqnarray}  
where $L_0$ is the kinetic term of relevant degrees of freedom,
$L_e$ is the graviton emission-absorption  term,  $L_r$ is rescattering term,
which is usually neglected \cite{Amati:1993tb}. The exact form of these 
terms is 
\begin{eqnarray}
L_0 =  -2\pd h^{--} \pd^{*} h^{++} + \pd_{+}{\pd^{*}}^2~~\phi
\pd_{-}\pd^2\phi^{*}. 
\end{eqnarray}  
Here $h^{--},$ $h^{++}$ are longitudinal degrees of freedom of the 
metric $h^{\mu\nu}.$ They have been defined 
by using the only non-zero components of the energy-momentum tensor
\begin{eqnarray*}
& T_{--} = kE\del(x^{-})\del^2({\bf{x}}), ~~~ 
T_{++} = kE\del(x^{+})\del^2({\bf{x-b}}), \\
& x^{\pm} = x^{0} \pm x^{3}, ~~~ {\bf{x}} = (x^1,x^2), ~~~ z = x^1 + ix^2, ~~~
\pd^{*} = {\pd}/{\pd{z^*}}, ~~~ k^2 = 8\pi G,
\end{eqnarray*} 
in the following way 
\begin{eqnarray*}
T_{\mu\nu}(x)h^{\mu\nu}(x) =  T_{--}(x)h^{--}(x) +  T_{++}(x)h^{++}(x),
\end{eqnarray*} 
where $x$ is four-dimensional coordinate. The $L_e$ is interaction 
Lagrangian. It is important for estimating two-loop correction. 
In order to estimate leading order correction  $L_e$ is not 
necessary. Equations of motion that come from the Lipatov action can 
be factorized  \cite{Amati:1993tb} by the substitution 
\begin{eqnarray*}
& h^{++}(x) = kE\del(x^{-})a(z,z^*), ~~~  
h^{--}(x) = kE\del(x^{+})\bar{a}(z,z^*) \\
& \phi = \frac{1}{2}k(kE)^2\frac{1}{2}\t(x^{+}x^{-})\varphi(z,z^*) \\
& \bar{a}(b-z,b-z^*) = a(z,z^*).
\end{eqnarray*} 

The physical picture depends on how large is impact parameter 
$b = |\bf{b}|$ in comparison 
with the effective Schwartzschild radius $R_S.$ If $b$ is much larger than 
$R_S$ the black hole does not appear and one has pure gravitational scattering 
process \cite{Amati:1993tb}. In the limit case $b >> R_S,$ the effective 
equations of motion are free field solutions which are Aichelburg-Sexl
shock waves \cite{'tHooft:rb},  
\begin{eqnarray}
a(z,z^*) = - \frac{1}{2\pi}\log  \frac{zz^*}{L^2},~~~\varphi = 0, \label{eqs}
\end{eqnarray}
where $L$ is a large-distance scale, which plays the role of infrared 
cutoff. Moreover, the Lipatov action has reproduced in the next approximation
with finite impact parameter $b$ the result of direct calculations of Ref. 
\cite{Amati:1990xe} of the eikonal amplitude up to two-loop level 
\cite{Amati:1993tb}. The value of the effective action on the effective 
equations of motion (\ref{eqs}) is \cite{Amati:1993tb}
\begin{eqnarray*}
& \dis{A(a,\bar{a},\varphi) = 2\pi Gs \le \bar{a}(0) + a(b) + 
\int d^2x\left[\bar{a}\pd^*\pd a + a\pd^*\pd\bar{a}\right]\ri} \\
& \dis{= 2\pi Gs \frac{1}{2} \le \bar{a}(0) + a(b)\ri = -2Gs\log\frac{b}{L}},
\end{eqnarray*}
which is phase of eikonal $\del(s,b).$ Then the leading eikonal function 
is equal to \cite{Amati:1993tb}  
\begin{eqnarray}
S = \exp\le-2\frac{i}{\hbar}Gs\log\frac{b}{L}\ri. \label{eikonal}
\end{eqnarray} 
Then by going from $s,b$ representation to $s,t$ representation one can write 
{\em two particles $\rar$ two particles} process scattering amplitude 
\begin{eqnarray}
& \dis{a(s,b) = e^{i\del(s,b)}}, \no\\
& \dis{ a(s,t) = \frac{1}{(2\pi)^2} \int~d^2x~e^{iq_ix^i} 
e^{i\del(s,b)}} \no\\
& = \dis{\frac{1}{2\pi}\int_0^{\infty}dbbJ_{0}(bq) e^{i\del(s,b)}}. \label{st}
\end{eqnarray}
Taking into account (\ref{eikonal}), we can reproduce the 't Hooft result   
\cite{'tHooft:rb} for the amplitude, 
\begin{eqnarray*}
\dis{a(s,t) \sim G \frac{s}{t}}.
\end{eqnarray*}
However, this is only scattering amplitude of two particles in Minkowski
background geometry. When the scattering parameter is small, the black hole
formation can occur. This case has been investigated in Refs. 
 \cite{D'Eath:hb,D'Eath:hd,D'Eath:qu} for the case of axisymmetric 
($b = 0$) two black holes collision. The form of the 
trapped surface which appears in such a collision has been found there. The 
criteria of the black hole formation is appearance of this trapped 
surface \cite{HAWELL}. The form of the trapped surface in four-dimensional 
collisions of two particles for $b \ne 0$ has been found in Ref. 
\cite{Eardley:2002re}. It has been found there that the correction changes
form of the geometrical cross-section a little by some factor of the order 
of 1.

\section{Graviton scattering in RS2 case}

The same idea can be applied to five-dimensional  case. Scattering 
amplitudes can be considered as value of the effective action calculated on the 
solution to the equations of motion. Here we have 
Randall-Sundrum model with infinite fifth dimension (RS2 scenario).
Emparan \cite{Emparan:2001ce} has extended the analysis of the shock 
wave of an ultrahigh-energy particle to scenarios with extra dimensions. The 
metric of shock wave in five dimensions has the form  
\begin{eqnarray*}
ds^2 = dy^2 + e^{-2y/R}(-dudv + dx^idx^i + h_{uu}(u,x_i,y)du^2), 
\end{eqnarray*}
where $u,v$ are light-cone coordinates and $x_i$ are coordinates in the 
directions transverse to propagation, $y$ is the coordinate in fifth dimension.
The equations of motion for the shock wave addition to AdS space $h_{uu}$ has 
been solved analytically
\cite{Emparan:2001ce} in the linearized approximation and then it has 
been argued 
that the solution to the linearized equation is in fact exact in this 
model \cite{Emparan:2001ce}. The analysis analogous to the four-dimensional 
case can be applied again and one obtains that in case $b > R$ the eikonal 
phase is 
\begin{eqnarray*}
\dis{\del(s,b) = -G_4s\le\log\frac{b}{R} - \frac{R^2}{2b^2}\ri}.
\end{eqnarray*}
The second term is the Kaluza-Klein  correction to the pure 
four-dimensional eikonal term, $G_4$ is four-dimensional coupling of 
gravity.  

In the limit $r<<R$ the solution for $h_{uu}$ is \cite{Emparan:2001ce}
\begin{eqnarray*}
\dis{h_{uu} = -4G_4 p\del (u)\left[-\frac{R}{r} + 
\frac{3}{2}\log(r/R) + \frac{3r}{8R} 
+ \dots \right],}
\end{eqnarray*} 
and this means that eikonal scattering phase is 
\begin{eqnarray*}
\dis{\del(s,b) = \frac{G_4sR}{2b}.}
\end{eqnarray*} 
In the $s,t$ representation (\ref{st}) the amplitude is  \cite{Emparan:2001ce}
\begin{eqnarray*}
a(s,t)  \sim G_4R \frac{s^2}{\sqrt{-t}}.
\end{eqnarray*}
This is the case of scattering. Again, as in the four dimensional case one 
has to search for the trapped surface to trace the black hole formation. 
At present, the result for trapped surface are absent in literature. 
Some estimations of the shape of horizon for the point static source on 
the brane have been done by Giddings \cite{Giddings}.

\section{Giddings' estimation for FB in gravity  with extra dimensions}

As is suggested, there is a minimum of the effective 
action $\Gamma$ in the form of AdS metric  
\begin{eqnarray*} 
ds^2 = dy^2 + e^{-2y/R}ds_M^2,
\end{eqnarray*} 
where $ds_M^2$ is Minkowski interval and that 
equations of motion of two particles in this background are non-linearized 
Einstein equations. One can consider static source of perturbations from 
the background metric on the brane as product of collision.   
In linearized approximation we just follow by the idea of Giddings  
when estimating 
size of horizon by the requirement  $h_{00} \sim -1.$ The field 
describing the perturbation about background is related to the variable 
of the effective action whose derivatives with respect to that variable 
are vertices of gravitons in this background.

The metric of the theory in extra dimension is taken in the form of 
perturbed AdS metric. It can be considered as some solution in low energy 
action of string theory. The perturbed AdS metric takes the form 
\cite{Giddings} 
\begin{eqnarray*} 
ds^2 = \le 1+h_{yy}\ri dy^2 + 
e^{-2y/R}\le\eta_{\mu\nu} + h_{\mu\nu}\ri{dx^\mu}{dx^\nu}
\end{eqnarray*} 
where $x$ are coordinates of our four dimensional world and $y$ is a 
coordinate in fifth extra dimension. The energy-momentum tensor is a source 
localized on the IR brane,
\begin{eqnarray*} 
T_{\mu\nu}(x,y) = S_{\mu\nu}(x)\del(y), ~~~ T_{yy} = T_{y\mu} = 0.
\end{eqnarray*} 
The point mass source considered in \cite{Giddings} is
\begin{eqnarray*} 
S_{\mu\nu}(x) = 2m\del^{d-1}(x)\del_{\mu}^0\del_{\nu}^0.
\end{eqnarray*} 
Linearized equations of motion can be solved by using Neumann Green's 
function 
\begin{eqnarray*} 
\Box\Delta_{d+1}(x,y;x',y') = \frac{\del^d(x-x')\del(y-y')}{\sqrt{-G}}.  
\end{eqnarray*} 
Under  Neumann boundary conditions, the solution takes the form 
\begin{eqnarray*} 
& \dis{h_{\mu\nu}(x,y) = 
-\frac{1}{2M_P^{d-1}}\int~d^dx'\sqrt{-g}\Delta_{d+1}(x,y;x',0)
\left[S_{\mu\nu}(x') - \eta_{\mu\nu}\frac{S_\lambda^\lambda(x')}{d-1} \right.} \no\\  
& + \dis{\left.\frac{\pd_\mu\pd_\nu}{\pd^2}\frac{S_\lambda^\lambda(x')}{d-1}\right].} 
\end{eqnarray*} 
This behaviour of the metric fluctuations $h_{\mu\nu}$ significantly 
simplifies at the large distances in four-dimensional space. 
The most interesting component  $h_{00}$ has the following 
behaviour
\begin{eqnarray*} 
h_{00}(x,y) = -\frac{km}{RM_P^{d-1}}\frac{e^{-M_1r+yd/R}}
{r^{d-3}}.   
\end{eqnarray*}
Here $k$ is a numerical constant and $M_1 = C/R,$ where $C$ is some
number. This component of the metric is important since it is responsible 
for the horizon formation, when $h_{00} = -1.$ Taking logarithm of the both 
parts, we come to the equation 
\begin{eqnarray*} 
M_1r -yd/R = \ln\left[\frac{km}{RM_P^{d-1}r^{d-3}}\right].
\end{eqnarray*} 
By solving this equation, one comes to the estimation of the area of the 
horizon in this model,
\begin{eqnarray*} 
r_h(m) \sim \frac{1}{M_1} \ln\left[\frac{kmM_1^{d-3}}{RM_P^{d-1}}\right].
\end{eqnarray*} 
Thus, the radius of horizon has logarithmic behavior in this brane model,
in complete correspondence to the Froissart bound \cite{Giddings}.

\section{Conclusion}

In this paper we have analysed general structure of the effective 
action for quantum gravity. We have shown that the effective action has 
the dressed effective fields as actual variables of the effective action. 
Kernels of the effective action written in terms of the dressed effective 
fields are constituent parts of the 
scattering amplitudes. In the eikonal approximation the scattering amplitudes 
can be found by solving the equations of motion derived from the 
Lipatov effective action.

\vskip 3mm
\noindent {\large{\bf{Acknowledgments}}}
\vskip 3mm

Work of I.K. is supported by Ministry of Education (Chile) under grant 
Mecesup FSM9901, by DGIP UTFSM, and by  Fondecyt (Chile)  
grants \#8000017 and \#1040368. I.K. is  grateful to Department of 
Physics of Brown University where this work has been started for 
the kind hospitality and financial support during his stay.

\end{document}